\documentclass[aps,prb,twocolumn,superscriptaddress]{revtex4}
\usepackage{amsmath}
\usepackage{epsfig}
\usepackage{graphics}

\newcommand{\DT}{{\cal D}_T}
\newcommand{\T}{{\cal T}}
\newcommand{\DL}{{\cal D}_L}

\def\ba{\begin{array}}
\def\ea{\end{array}}

\begin{document}

\title{Supercurrent-induced Peltier-like effect in
superconductor/normal-metal weak links} 
\author{Tero T. Heikkil\"a}
\email[Email address:]{Tero.T.Heikkila@hut.fi}
\thanks{We thank N. Birge for the idea to this work and J.\ von Delft,
  D. J. van Harlingen and J. Pekola for suggestions on the
  manuscript. This work was supported by the Graduate School in
  Technical Physics at the Helsinki University of Technology.}

\affiliation{Materials Physics Laboratory, Helsinki University of
Technology, P.O. Box 2200, FIN-02015 HUT, Finland}
\affiliation{Low Temperature Laboratory, Helsinki University of
Technology, P.O. Box 2200, FIN-02015 HUT, Finland}

\author{Tommy V\"ansk\"a}

\affiliation{Materials Physics Laboratory, Helsinki University of
Technology, P.O. Box 2200, FIN-02015 HUT, Finland}

\author{Frank K. Wilhelm}
\affiliation{Sektion Physik and CeNS, Ludwig-Maximilians-Universit\"at, 
Theresienstr. 37, D-80333 M\"unchen, Germany}

\date{\today}

\begin{abstract}
The local nonequilibrium quasiparticle distribution function in
a normal-metal wire depends on the applied voltage over the
wire and the type and strength of different scattering mechanisms. We
show that in a setup with superconducting reservoirs, in which the 
supercurrent and the dissipative current flow (anti)parallel, 
the distribution
function can also be tuned by 
applying a supercurrent between the contacts. Unlike the usual control 
by voltage or temperature, this leads 
to a Peltier-like effect: the supercurrent converts an externally
applied voltage into a 
difference in the effective
temperature between two parts of the system maintained at the same potential. 
We suggest an experimental setup for probing this phenomenon and mapping out
the controlled distribution function. 
\end{abstract}

\pacs{74.25.Fy,74.40.+k,74.50.+r,74.80.Fp}

\maketitle

Many of the well-understood phenomena in mesoscopic physics can be probed
within the linear response of a physical
system to an applied external perturbation, i.e., they are 
governed by equilibrium physics. 
Recently the attention has turned more towards the
study of effects far from equilibrium. The quasiparticle
distribution function $f(x;E)$ characterizing the nonequilibrium was
measured in a normal-metal (N) wire between two large reservoirs
\cite{pothier97,pierre01} through a superconducting (S) tunnel probe. This
yielded useful information on the residual interactions between the
Fermi-liquid quasiparticles. This nonequilibrium distribution was used
to control the supercurrent in a normal-metal weak 
link. \cite{baselmansnature,baselmans01,huang} Both of these setups
serve as different types of local probes for $f(x;E)$.

As a further step, we describe the control of $f(x;E)$ via
the supercurrent. We show that, unlike other control parameters, it
changes the profile of the effective
temperature through the sample in the form of a large Peltier effect,
i.e., heating the electrons in one part of the structure, and cooling
them in another --- even in the case of complete electron-hole
symmetry. Moreover, we show how the two types of measurements for
$f(x;E)$ can be combined within the same sample. 

We concentrate on studying a diffusive normal-metal wire, where elastic
scattering is the dominant scattering mechanism. In the absence of 
superconductivity and for wires much shorter than the inelastic
scattering length, the steady-state distribution
function between two reservoirs with 
chemical potentials $\mu_1$ and $\mu_2$ has a double-step form,
interpolating between the two Fermi functions in the reservoirs.
\cite{pothier97}

When the N reservoirs are replaced by
superconducting ones, the leading transport mechanism at energies 
below the superconducting gap $\Delta$ is Andreev reflection. \cite{andreev64}
This leads to 
a penetration of superconducting correlations into the
N wire (superconducting proximity effect). It modifies the
charge and energy conductivities and we may introduce the corresponding
diffusion coefficients ${\cal D}_T(x;E)$ and ${\cal D}_L(x;E)$
depending on space and energy. \cite{karlsruhereview} More
importantly, the proximity effect allows 
supercurrents to flow through the N wire. To describe these 
effects, it is convenient to separate $f(x;E)$ into symmetric and
antisymmetric parts relative to the chemical potential $\mu_S$ of the
superconductor,   
\begin{align}
f_T(E)&\equiv 1-f(\mu_S-E)-f(\mu_S+E),\\ f_L(E) &\equiv
f(\mu_S-E)-f(\mu_S+E).
\end{align}
Below, we choose $\mu_S=0$. These functions describe charge and energy
distributions, respectively. 
They satisfy the kinetic equations \cite{karlsruhereview,galaktionov}
\begin{align}
\frac{\partial j_T}{\partial x}=0,\quad j_T \equiv \DT(x) \frac{\partial
f_T}{\partial x} + j_E f_L + \T(x) \partial_x f_L;
\label{eq:chargecurrent}\\ 
\frac{\partial j_L}{\partial x}=0,\quad j_L \equiv \DL(x) \frac{\partial
f_L}{\partial x} + j_E f_T - \T(x) \partial_x f_T.
\label{eq:energycurrent}
\end{align}
Here we assume no energy relaxation, so the kinetic equations
describe the conservation of $j_T(E)$ and $Ej_L$, the 
spectral charge and energy currents, respectively. Terms $\DT$, $\DL$,
$j_E$, and $\T$ can be found from quasiclassical equations for the 
retarded Green's function in the diffusive limit.
\cite{usadel,karlsruhereview} All of them depend on the phase
difference $\phi$ between the superconductors such that for $\phi=0$,
$j_E$ and $\T$ vanish. In our case, the charge diffusion coefficient
$\DT$ is increased at most up to 20\% from its normal-state value $\DT=1$,
\cite{charlat96} whereas for energies below $\Delta$,
$\DL$ tends towards zero near the S
interface, effectively prohibiting energy transport. The term
$\T(x;E,\phi)$ (Ref.~\onlinecite{galaktionov}) is obtained
as a cross term from the retarded and advanced Green's functions.
In general, it is much smaller
than the other coefficients.
The supercurrent is 
described by a spectrum $j_E(E;\phi)$ of supercurrent-carrying
states, \cite{wilhelmprl,yip,heikkila02} which yields a contribution
$j_E f_L(x)$ to the spectral charge current and, under nonequilibrium
conditions involving $f_T(x) \neq 0$, a contribution to the energy
current $Ej_E f_T(x)$. 

These kinetic equations have to be supplied with boundary conditions.
At N reservoirs, electrons are simply transmitted and the distributions
have to match Fermi functions with shifted chemical potentials.
At the NS interface for $|E|<\Delta$, Andreev reflection 
prohibits the
transfer of energy into S yielding $j_L=0$. The charge distribution is
continuous, which 
leads to $f_T(E) = 0$ at the NS interface assuming that there is no 
charge imbalance in the leads.

The nonequilibrium distribution function may be characterized through
its moments, 
the local chemical potential 
$\mu(x)$ and the local effective temperature $T_{\rm eff}(x)$. The
previous characterizes the charge distribution function as
$\mu(x)=\int_0^\infty dE\;f_T(x;E)$. The effective
temperature describes the amount of heat in the electron system and
is related to the energy distribution function via
\begin{equation}
\frac{e^2{\cal L}_0}{2} T_{\rm eff}^2(x) = \int_0^\infty dE
E[f_{L,0}(x;E)-f_L(x;E)], 
\end{equation}
where ${\cal L}_0=(\pi^2k_B^2/3e^2)$ is the Lorenz number and the
corresponding zero-temperature distribution has a step-function form
$f_{L,0}(x;E \ge 0) = 1-\theta[E-\mu(x)]$.

In the absence of the supercurrent, the kinetic Eqs.
(\ref{eq:chargecurrent}) and (\ref{eq:energycurrent}) are not coupled
and, consequently, there is no 
thermoelectric coupling between the applied voltage and the energy currents. 
This results from the
assumption of bands with complete electron-hole symmetry in the derivation 
of the formalism. Beyond the limits of the formalism, it is known
that  
electron-hole symmetry breaking leads to small thermoelectric effects
in normal metals, limited by the tiny factor $k_B T/\epsilon_F$.
\cite{ashcroftmermin}

\begin{figure}
\centering
\includegraphics[width=0.77\columnwidth,clip]{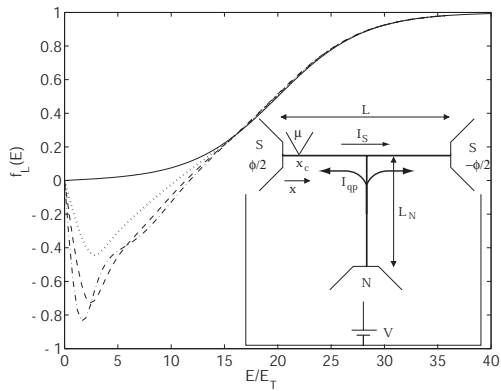}
\setlength{\unitlength}{1cm}
\caption{Supercurrent-driven distribution function $f_L(E)$
at the left NS interface as a function of energy for $\phi=0$ (solid),
$\phi=0.12 \pi$ (dotted), $\phi=0.24 \pi$ (dashed), and $\phi=\pi/2$
(dash dotted). The result is obtained with $T=4E_T/k_B$ and $V=20
E_T/e$. The corresponding changes of $f_L$ by the supercurrent in the
right arm have the opposite sign. Inset: the system under
consideration. We assume symmetric horizontal wires of length
$L/2$. This length defines the Thouless energy of the weak link,
$E_T\equiv \hbar D/L^2$. The resistance of the weak link is $R_{\rm
  w}$ and of the vertical wire $R_{\rm c}$. Measurement of the
predicted effects can be performed by placing a superconducting tunnel
probe at position $x=x_c$, near the NS interface.}  
\label{fig:flphasedep}
\end{figure}

Below, we study a multiterminal
setup depicted in the inset of Fig.\ \ref{fig:flphasedep}: varying the
voltage between the N and S reservoirs while maintaining the
superconductors at equal potentials allows one to vary the
distribution function in the phase-coherent wire. Such a device has
already been implemented for controlling the critical current for the
dc Josephson effect. \cite{huang} 
It permits
to study
the 
supercurrent under nonequilibrium conditions
without the
complications caused by the ac Josephson effect and is, hence, an appropriate
system for demonstrating the physics outlined above: 
As the energy flow $Ej_Ef_T(x)$ carried by the
extra quasiparticles injected into the supercurrent-carrying states
cannot pass into the superconductors, it has to be counterbalanced by
another energy flow. This flow is driven by the gradient of the energy
distribution function $E\DL \partial_x f_L$ and hence, the applied
control voltage is converted into a gradient of the effective temperature
through the supercurrent.

\begin{figure}
\centering
\includegraphics[width=0.9\columnwidth,clip]{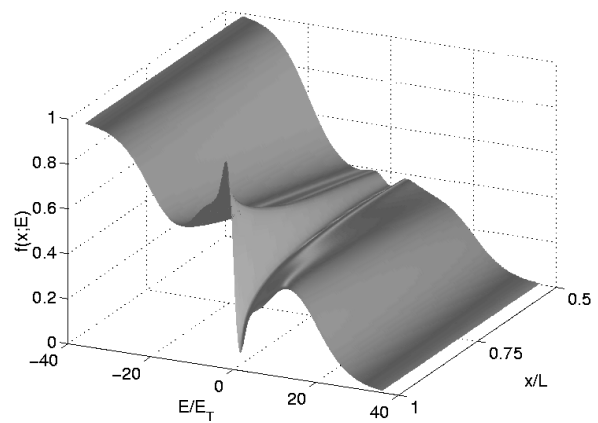}
\caption{Quasiparticle distribution function $f(x;E)$ in the right-horizontal arm for voltage $V=20E_T/e$, temperature $T=4E_T/k_B$, and 
phase difference $\phi=\pi/2$ between the superconductors. The large
deviations from the rounded staircase form are created by the
supercurrent flowing in the structure.}
\label{fig:distfuncs3d}
\end{figure}

Solving Eqs.\ (\ref{eq:chargecurrent}) and (\ref{eq:energycurrent}) for
$\phi=0$ and $E<\Delta$ is similar to a two-probe
N-S case \cite{karlsruhereview}: $f_L$
stays constant throughout the phase-coherent wire at its value in the
N reservoir, $f_L^0(V)=(\tanh[(E+eV)/2k_BT]+\tanh[(E-eV)/2k_BT])/2$
and $f_T$ is slightly modified from the linear space dependence due to the
proximity effect on $\DT$. \cite{charlat96} Increasing the phase
$\phi$ induces a finite supercurrent into the weak link, thereby
coupling $f_L$ and $f_T$. First neglecting the small
coefficient $\T$, we get
\begin{align}
\frac{\partial f_L}{\partial x}=-j_E\frac{f_T}{\DL},\quad
\frac{\partial}{\partial x}\left(\DT\frac{\partial f_T}{\partial x}\right)
= j_E^2 \frac{f_T}{\DL}.
\end{align}
Assuming that $j_E$ is small, we observe that the major change due to the
supercurrent is expected for $f_L(E,x)$; particularly, it will depend
on space.

In general, a closed-form solution for
$f_L(x;E)$, $f_T(x;E)$ cannot be found. Therefore, we focus on a
numerical solution of both the spectral and kinetic equations. Here
and below, we assume that all the energies are below $\Delta$. The
effect of the supercurrent on the distribution functions is clearest
at a low temperature $k_B T \lesssim E_T$. The resulting distribution
function $f(x,E)$ for the right-hand horizontal arm is plotted in
Fig.\ \ref{fig:distfuncs3d} for $\phi=\pi/2$, yielding a supercurrent
close to its maximum. As expected, the antisymmetric part of $f(x;E)$
has become space dependent, its energy dependence following that of
$j_E$.  Fixing a position in space, chosen, for example, near the NS
interface in the left-hand side horizontal arm, allows us to observe
how the distribution function changes as a function of phase $\phi$,
i.e., as it is driven by the supercurrent. This is illustrated in
Fig. \ref{fig:flphasedep}, where $f_L(E;\phi)$ is plotted for a few
values of $\phi$. 

In the three-probe case, the chemical potential $\mu(x)$ interpolates 
nearly linearily between the chemical potentials of the superconductor
and the normal reservoir and varies only little with the supercurrent. 
The changes in
the effective temperature $T_{\rm eff}$ are much more pronounced. In
the absence of the supercurrent, $T_{\rm eff}$ is
\begin{equation}
T_{\rm eff}^0\equiv T_{\rm
  eff}(\phi=0)=\sqrt{T^2+\{V^2-(\mu(x)/e)^2\}/{\cal L}_0}. 
\label{eq:t0nocurr}
\end{equation}
Both $T_{\rm eff}(x;\phi=0)$ and $\mu(x;\phi=0)$ are symmetric in the
two horizontal arms.
The supercurrent-induced change in $f_L(x;E)$ 
can be described through the change of the effective temperature compared to
Eq.\ (\ref{eq:t0nocurr}), such that $T_{\rm eff}(x)=[T_{\rm
  eff}^0(x)^2+S(x;V)+\delta \mu(x;\phi)]^{1/2}$, 
where  
\begin{equation}
S(x;V) = \frac{6}{\pi^2 k_B^2} \int_0^\infty dE
E(f_L^0(E;V)-f_L(x;E))
\end{equation}
and $\delta \mu(x;\phi) \equiv [\mu(x;\phi)^2-\mu(x;\phi=0)^2]/2$
describes the change in the local chemical potential due to the
supercurrent, a much smaller term than $S(x)$. The kinetic equations
imply that the supercurrent-induced change of
the distribution function $f_L$ is 
antisymmetric between the two arms, hence so is $S(x)$, i.e.,
$T_{\rm eff}$ increases in one arm and decreases in the other
one. Hence, the system works analogously to a Peltier device, where
the control current is replaced by the supercurrent: the supercurrent
``cools'' one part of the system, transferring the heat to another
part.  
\begin{figure}
\centering
\includegraphics[width=0.77\columnwidth,clip]{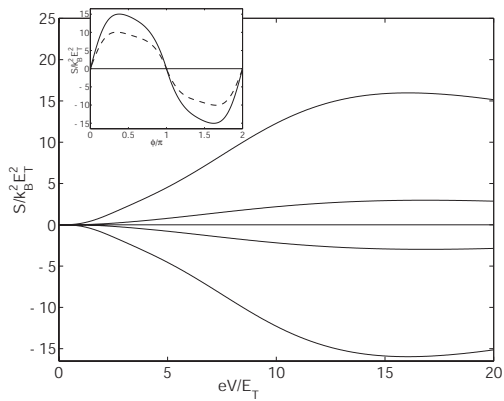}
\caption{Supercurrent-induced change $S(x;V)$ in the effective
temperature as a function of voltage $eV/E_T$ at different positions
in the weak link with $\phi=\pi/2$. From top to bottom: $x=0$,
$x=L/4$, $x=L/2$, $x=3L/4$, and $x=L$. Here $x=0$ and $x=L$ correspond to
the left and right S interfaces and $x=L/2$ to the crossing
point. Inset shows the phase dependence of $S(x=0)$ for $eV=12E_T$
(solid) and $eV=8E_T$ (dashed). In both curves, the bath temperature
$T=0$.}  
\label{fig:effTchange}
\end{figure}
The supercurrent-induced
temperature change $S(x;V)$ is illustrated in Fig.\
\ref{fig:effTchange}.

To obtain an estimate for $S(x;V,\phi)$, we approximate
$\DT(E;x) = 1$ and find
\begin{equation}
S=\frac{2E_T^2R_{\rm w}}{L^2(R_{\rm w}+R_{\rm c})}\int_0^\infty dE
E j_E(E) f_T^0(E) \int_x^{L/2} \frac{x'dx'}{\DL(x';E)}.
\end{equation}
At low temperatures $k_B T \ll eV$, $f_T$ 
reduces to a step
function around the potential $eV$ in the reservoir, 
cutting the integration off at
$E=eV$. Thus, this current-induced temperature change, which is
similar to the Peltier effect, is much larger than in conventional
single-metal setups.

For the measurement of the predicted effects in the distribution
function, we suggest a setup shown in the inset of
Fig. \ref{fig:flphasedep} -- very similar to those used in
Refs. \onlinecite{pothier97,pierre01}. Such a setup has also been used as a
local thermometer \cite{rowell76} of the electronic temperatures. A
superconducting wire is connected to the horizontal arm via a
highly resistive tunneling layer (I) at position, say, $x=x_c$. 
The dc current is then given by
the tunneling quasiparticle current
\begin{equation} 
I_J=\frac{1}{eR}\int_{-\infty}^\infty dE \rho_S(E+\mu)
\rho(E)[f_0(E+\mu)-f(E)], 
\end{equation}
where $N_S\rho_S(E)$ is the BCS density of states (DOS) of the tunnel
probe, $N_N\rho(E)$ is the local DOS in the mesoscopic wire at
$x=x_c$, $N_S$ and $N_N$ are the normal-state DOS's for the two
materials at $E=E_F$, $f_0(E)$ is the Fermi function, and $f(E)$ is the
distribution function to be measured. When all the wires are in the normal
state, the resistance through the tunnel junction is $R$. We can
separate this expression as  $I_J=I_1 + I_2$, where $I_1$ is the
tunneling current for the equilibrium system $V=0$, probing $\rho(E)$
and 
\begin{equation}
\begin{split}
I_2&=\frac{1}{eR}\int_0^\infty {\rm d}E \rho(E)
\big\{f_T(E)[\rho_S(\mu+E)+\rho_S(\mu-E)]\\
+&\left[f_L(E)-\tanh(E/2k_BT)\right][\rho_S(\mu+E)-\rho_S(\mu-E)]\big\} 
\label{eq:i2}
\end{split}
\end{equation}
depends on the state of the wire, and for an equilibrium state, $V=0$, 
vanishes. 
In order
to isolate $I_2$, 
one can first determine $I_1$ as a function of the supercurrent
  by investigating
the equilibrium case.
Then,
$I_1$ may be substracted from the nonequilibrium results,
leaving only currents $I_2$. 
Moreover, $I_2(\mu)+I_2(-\mu)$ is proportional to the first part of
Eq.\ (\ref{eq:i2}), dependent on $f_T(E)$, and $I_2(\mu)-I_2(-\mu)$ to
the second, dependent on $f_L(E)$. 

With the above setup, the distribution functions may be characterized
as a function of both the voltage $V$ and the supercurrent driven
through the weak link. The
setup also makes it possible to measure the local 
distribution function both through the NIS contact and through the
SNS critical current. These two independent probes should permit
to distinguish the contributions from different inelastic
scattering effects along the lines of Ref.~\onlinecite{pothier97}.

So far we have completely neglected inelastic scattering in the wires.
We can
include energy relaxation due to electron-electron scattering 
phenomenologically, generalizing the method of 
 Ref.~\onlinecite{nagaev95} to include the
effect of supercurrent. In the limit $L \gg l_\varepsilon$,
we may describe the nonequilibrium distribution functions by Fermi
functions with local chemical potential and temperature. In this case,
assuming for simplicity $\DT = 1$ and $\T = 0$, we 
can 
integrate the two kinetic equations over energy,
obtain kinetic equations for
$\mu(x)$ and $T_{\rm eff}(x)$ and
find in the limit of high $\Delta$
\begin{align}
\partial_x^2 \mu(x) &= - \partial_x I_S(x),\\
e^2{\cal L}_0 \tilde{T}(x) \partial_x T_{\rm eff}(x) &=
-\tilde{\mu}(x) \partial_x \mu(x) + Q_S.
\end{align}
Here $I_S(x) = [\int dE j_E(E) f_L(E,x)]/2$ is the local
supercurrent, $2{\cal L}_0e^2\tilde{T}=-\int dE E \DL \partial_T f_L$
and $2\tilde{\mu}(x)=-\int {\rm d}E E \DL \partial_\mu f_L$ describe the
local temperature and chemical potential modified by $\DL$,
respectively, and $Q_S = [\int dE E j_E(E) f_T(E,x)]/2$ is the energy
current carried by the nonequilibrium supercurrent. The first equation
states the conservation of the total current whereas the latter
describes the temperature profile. In the absence of the proximity
effect, these yield the effective temperature given in 
Eq.~(\ref{eq:t0nocurr}). Similarly as above, the effective temperature
can also in this case be tuned via the supercurrent, through the
control of $Q_S$. 

The predicted effect resembles a previously studied phenomenon in
bulk superconductors, \cite{thermalchargeimbalance} where a
temperature gradient along with a
supercurrent generates a  charge imbalance in S.
Here, the finite voltage (described through $f_T$) along with the
supercurrent produces a temperature gradient (spatial variation of $f_L$).
\cite{charimnote}

In Ref.~\onlinecite{venkat}, another thermoelectric effect, the thermopower,
has been measured experimentally in a similar type of a system. The
coupling of the distribution functions through the supercurrent may
explain part of the observed effects. In Ref.~\onlinecite{seviouremf},
thermopower has been studied in the regime of high tunnel barriers and
within  linear response, leading also to an unexpectedly large
effect. In that paper, all the distribution
functions are, besides minor corrections, in quasiequilibrium: the
transport is essentially driven by the discontinuities at the
tunneling barriers. Moreover, Ref.~\onlinecite{claughtonlambert}
studies the Andreev interferometers through a numerical scattering
approach, and predicts an oscillating thermopower as a function of the
phase. However, there the quasiparticle current and supercurrent do
not flow in parallel and the magnitude of the effect may be
strongly affected by the very small size of the studied structure.

Summarizing, we predict that in a nonequilibrium situation created by
applying a voltage between a normal metal and two superconductors, the
nonequilibrium distribution functions in the normal-metal wire can be
tuned by the supercurrent flowing between the superconductors. This
results in a supercurrent-controlled Peltier effect. The predicted
effect can be observed by the measurement of the tunneling current
from an additional superconductor.


\begin{thebibliography}{21}
\bibitem{pothier97} H. Pothier, {\it et al.}, Phys. Rev. Lett. {\bf
79}, 3490 (1997).
\bibitem{pierre01} F. Pierre, {\it et al.} Phys. Rev. Lett. {\bf 86},
1078 (2001).
\bibitem{baselmansnature} J. J. A. Baselmans, A. F. Morpurgo, B. J. van
Wees, and T. M. Klapwijk, Nature (London) {\bf 397}, 43 (1999).
\bibitem{baselmans01} J. J. A. Baselmans, B. J. van Wees, and
  T. M. Klapwijk, Phys. Rev. B {\bf 63}, 094504 (2001).
\bibitem{huang} J. Huang {\it et al.}, Phys. Rev. B {\bf 66},
  020507(R) (2002).
\bibitem{andreev64} A. F. Andreev, Zh. Eksp. Teor. Fiz. {\bf 46}, 1823
(1964) [JETP {\bf 19}, 1228 (1964)].
\bibitem{karlsruhereview} W. Belzig {\it et al.}, Superlattices
Microstruct. {\bf 25}, 1251 (1999).
\bibitem{galaktionov} S.-W. Lee, A. V. Galaktionov, and C.-M. Ryu,
  J. Korean Phys. Soc. {\bf 34}, S193 (1999).
\bibitem{usadel} K. D. Usadel, Phys. Rev. Lett. {\bf 25}, 507 (1970).
\bibitem{charlat96} P. Charlat {\it et al.}, Phys. Rev. Lett. {\bf 77},
4950 (1996). This modification of $\DT$ can also be tuned by a phase
between two superconductors in an Andreev interferometer [P. G. N. de
  Vegvar, T. A. Fulton, W. H. Mallison, and R. E. Miller,
  {\it ibid.} {\bf 73}, 1416 (1994)]. This tuning leads to
the phase dependence of $\mu(x;\varphi)$.
\bibitem{wilhelmprl} F. K. Wilhelm, G. Sch\"on, and A. D. Zaikin,
  Phys. Rev. Lett. {\bf 81}, 1682 (1998).
\bibitem{yip} S. K. Yip, Phys. Rev. B {\bf 58}, 5803 (1998).
\bibitem{heikkila02} T. T. Heikkil\"a, J. S\"arkk\"a, and
F. K. Wilhelm, Phys. Rev. B {\bf 66}, 184513 (2002).
\bibitem{ashcroftmermin} N. W. Ashcroft and N. D. Mermin, {\it Solid
  State Physics} (Saunders College, Orlando, 1976). 
\bibitem{rowell76} J. M. Rowell and D. C. Tsui, Phys. Rev. B {\bf 14},
  2456 (1976). 
\bibitem{nagaev95} K. E. Nagaev, Phys. Rev. B {\bf 52}, 4740 (1995).
\bibitem{thermalchargeimbalance} C. J. Pethick and H. Smith,
Phys.\ Rev.\ Lett.\ {\bf 43}, 640 (1979); J. Clarke, B. R. Fjordb\o ge, 
and P. E. Lindelof, {\it ibid.} {\bf 43}, 642 (1979);
A. Schmid and G. Sch\"on, {\it ibid.} {\bf 43}, 793 (1979);
J. Clarke and M. Tinkham, {\it ibid.} {\bf 44}, 106 (1980). 
\bibitem{charimnote} However, the mechanism for the coupling between
  the temperature gradient and charge imbalance is different in these
  two cases: whereas in bulk superconductors, it arises from the 
superconducting condensate that converts quasiparticle currents into
supercurrents, the phenomenon depicted here is due to the energy
current carried by the supercurrent in the nonequilibrium state.
\bibitem{venkat} J.\ Eom, C.J.\ Chien, and V.\ Chandrasekhar, Phys.\ 
Rev.\ Lett.\ {\bf 81}, 437 (1998). 
\bibitem{seviouremf} R.\ Seviour and A.\ F.\ Volkov, Phys. Rev. B {\bf 62},
R6116 (2000). 
\bibitem{claughtonlambert} N. R. Claughton and C. J. Lambert,
  Phys. Rev. B {\bf 53}, 6605 (1996). 
\end{thebibliography}
\end{document}